\documentclass[aps,superscriptaddress,twocolumn]{revtex4-2}
\usepackage{graphicx}  
\usepackage{dcolumn}   
\usepackage{bm}        
\usepackage{amssymb}   
\usepackage{amsmath}
\usepackage{amsfonts}
\usepackage[colorlinks=true,linkcolor=red, citecolor=red]{hyperref}
\usepackage{tikz}
\usetikzlibrary{shapes,arrows}
\usepackage{siunitx}
\usepackage{physics}
\usepackage{hhline} 
\usepackage{booktabs}
\usepackage{caption}
\usepackage{subcaption}

\begin{document}

\title{Joule Thomson expansion of $\kappa$-deformed Schwarzschild-AdS black hole}

\author{Vishnu Rajagopal}
\email{vishnu@hunnu.edu.cn}
\affiliation{Department of Physics and Synergetic Innovation\\
Center for Quantum Effects and Applications,\\
Hunan Normal University, Changsha, Hunan 410081, China}
\affiliation{Institute of Interdisciplinary Studies,\\
Hunan Normal University, Changsha, Hunan 410081, China}

\author{Puxun Wu}
\email{pxwu@hunnu.edu.cn}
\affiliation{Department of Physics and Synergetic Innovation\\
Center for Quantum Effects and Applications,\\
Hunan Normal University, Changsha, Hunan 410081, China}
\affiliation{Institute of Interdisciplinary Studies,\\
Hunan Normal University, Changsha, Hunan 410081, China}

\begin{abstract}

We investigate the Joule-Thomson expansion of a non-commutative Schwarzschild-AdS black hole, whose space-time geometry is described by the Lie-algebraic $\kappa$-deformed space-time, and show that the $\kappa$-deformation parameter induces Joule-Thomson expansion in an uncharged-AdS black hole. We find the ratio of minimum inversion temperature to critical temperature $T_i^{(min)}/T_c \simeq 0.493$, is independent of the deformation parameter, and is in close agreement with the value for Reissner-Nordstrom black hole. Furthermore, we show the existence of a minimum black hole mass (depending on the $\kappa$-deformation parameter) below which the Joule-Thomson expansion vanishes.

\end{abstract}

\maketitle

\section{Introduction}

The thermodynamic description of black holes (BH) plays an important role in establishing deep interconnections between the general relativity and quantum mechanics.
The seminal contributions of Bekenstein and Hawking \cite{Bekenstein:1972tm,Bekenstein:1973ur,Bardeen:1973gs,Bekenstein:1974ax,Hawking:1974rv,Hawking:1975vcx,Hawking:1976de}, has shown that BHs possess entropy proportional to their horizon area and radiate thermally, has offered a unique window to understand the nature of quantum gravity. The thermodynamic investigation of BHs has began with the pathbreaking work of Hawking and Page \cite{Hawking:1982dh}, which revealed the Hawking-Page phase transition between Schwarzschild-AdS BHs and thermal AdS space. Subsequently, the charged RN-AdS black BHs gas been shown to undergo first-order phase transitions, exhibiting a striking resemblance with the liquid-gas phase transition of van der Waals fluids \cite{Chamblin:1999tk,Chamblin:1999hg}. A major development has occurred when the cosmological constant and its conjugate quantity were interpreted as thermodynamic pressure and volume, respectively, and the black hole mass was identified as enthalpy. This has lead to the formulation of ﬁrst law of BH thermodynamics and the corresponding Smarr relations in the extended phase space \cite{Kastor:2009wy}. Within this framework, charged AdS BHs have been shown to exhibit $P-V$ criticality and phase transition behavior that coincide strikingly with the van der Waals system \cite{Kubiznak:2012wp}, and similar phase transitions have further been observed in rotating, higher-dimensional, and various other AdS BH configurations \cite{Dolan:2011xt,Gunasekaran:2012dq,Belhaj:2013cva,Cai:2013qga,Dehghani:2014caa,Zhang:2014uoa,Zhang:2015ova,Hennigar:2015esa,Li:2020xkh,Gogoi:2021syo,Cong:2021jgb,Astefanesei:2023sep}.

Different AdS BH configurations have been investigated rigorously to deepen the understanding of BH thermodynamics within the framework of various thermodynamic phenomenon \cite{Ditta:2024iky,Mustafa:2024mvx,Mustafa:2024kjy,Ditta:2025vsa,Ditta:2024lnb,Ashraf:2025nvt,Ashraf:2024itj,Ashraf:2024xwm,Ashraf:2025rdb,Ashraf:2025cnz,Saleem:2025vyq,Bouzenada:2025ypa,Ashraf:2025uhj,Maurya:2025jch,Ditta:2025kec,Ashraf:2025elg,Ditta:2025oiw,AbdelmalekBouzenada:2026zwu,Bouzenada:2025rsp,Ditta:2025vsa}. In this regard, the charged AdS BHs have been shown to exhibit an interesting thermodynamic phenomenon, namely the Joule-Thomson expansion process, which is characterized by an inversion temperature, whose corresponding inversion curve separates the isenthalpic curves in the $T-P$ plane into cooling and heating regions \cite{Okcu:2016tgt}. This Joule Thomson expansion process has also been observed in Kerr-AdS BHs \cite{Okcu:2017qgo} and has been generalised to different AdS BHs \cite{Lan:2018nnp,Zhao:2018kpz,Mo:2018rgq,Lan:2019kak,Cisterna:2018jqg,MahdavianYekta:2019dwf,Mo:2018qkt,Li:2019jcd,Pu:2019bxf,Bi:2020vcg,Barrientos:2022uit}. Recently the Joule-Thomson expansion has been studied extensively in the presence of cosmological fluids, such as quintessence and phantom dark energy \cite{Rajani:2020mdw,Yin:2021akt,MoraisGraca:2021ife,Javed:2024ohv,Ahmed:2025qza,Gogoi:2026ijd}.

The quantum corrected BH solutions and the associated BH thermodynamics have been explored extensively in the recent times and these quantum corrections are shown to induce some interesting features, such as phase-transition, criticality and Joule-Thomson expansion. Most importantly, such quantum corrected BHs are constructed within the framework some of quantum gravity approaches such as loop quantum gravity (LQG) and non-commutative (NC) space-time geometry. The NC space-time geometry effectively captures the quantum structure of space-time, and naturally incorporate the quantum gravitational effects through the fundamental length scale \cite{Snyder:1946qz,Doplicher:1994zv}. Remarkably, the minimal length scale (or NC parameter) of Moyal space-time \cite{Douglas:2001ba}, which is a canonical type NC space-time, has been shown to remove the curvature singularity of classical BHs through an regular de Sitter core, induced by NC parameter \cite{Nicolini:2005vd}. Similarly, the minimal length scale of $\kappa$-deformed NC space-time \cite{Arzano:2021hpg}, which is a Lie-algebraic type NC space-time, has been shown to remove the big-bang singularity through bounce behaviour in the recent times \cite{Rajagopal:2025qyp}. Such NC parameters are shown to modify the phase transition behaviour, critical ratio and the inversion temperatures of the standard BH thermodynamics \cite{Banerjee:2008gc,Modesto:2010rv,Liang:2017rng,Filho:2022zdh,Filho:2024zxx,Panja:2026hja}.

The NC parameter of Moyal space-time is shown to transform the Hawking-Page transition from a simple first-order phase transition into a van der Waal's-like phase diagram with a critical point, beyond which the transition becomes a smooth crossover \cite{Nicolini:2011dp}. Upon treating the NC parameter and its conjugate as thermodynamic variables in the extended phase, the NC parameters of both Moyal and $\kappa$-deformed space-times, are shown to to induce phase-transition and criticality in uncharged Schwarzschild-AdS BHs \cite{Wang:2024jlj,Tan:2024jkj,Wang:2025ycl,Wang:2025alf,Kumara:2026uwi}. Similarly, the LQG parameter of the quantum corrected Schwarzschild-AdS BH has also been shown to exhibit such phase transition and criticality behaviours \cite{Wang:2024jtp}.

Recently, the characteristic NC \cite{Wang:2024jlj,Tan:2024jkj,Wang:2025ycl,Wang:2025alf,Graca:2021ker} and LQG \cite{Wang:2024jtp} corrections to the Schwarzschild-AdS BH solutions, obtained in the frameworks of Moyal space-time and LQG, are shown to introduce Joule-Thomson effect in the uncharged and non-rotating AdS-BHs. Moreover, the ratio of minimum inversion temperature to critical temperature is shown to be independent of NC and LQG parameters (or minimal length scale). In this context it is important to ask the fundamental question whether this Joule-Thomson effect is a universal behaviour exhibited by all the quantum gravity corrected BHs or is this behaviour visible only among the Moyal and LQG corrected BHs? In order to address this question, it is necessary to investigate whether the $\kappa$-deformed NC corrections to the Schwarzschild-AdS BH solution exhibit the Joule-Thomson expansion or not. So in this work, we use the $\kappa$-deformed Schwarzschild-AdS BH solution obtained in \cite{Kumara:2026uwi} to investigate the Joule-Thomson effect.

This paper is organised in the following way. In Sec.(\ref{sec2}), we provide a brief discussion on the thermodynamics of $\kappa$-deformed Schwarzschild-AdS BH. Further, we compute the pressure, temperature and volume at critical point and evaluate the corresponding critical ratio. In Sec.(\ref{sec3}), we study the Joule-Thomson expansion in detail, and also evaluate the ratio of minimum inversion temperature to critical temperature. In Sec.(\ref{sec4}), we summarise our results and provide the concluding remarks.


\section{$\kappa$-deformed Schwarzschild BH thermodynamics}\label{sec2}

The minimal length scales corrections of the NC space-time structures \cite{Nicolini:2005vd} and the T-string duality \cite{Nicolini:2019irw} are shown to deform the point-like matter distributions and introduces quantum corrections to the energy-momentum tensor, which is shown to generate minimal length scale induced quantum corrections to the BH solutions, through the Einstein's field equation. In this regard, the $\kappa$-deformed Schwarzschild-AdS black hole solution has also been constructed rigorously in \cite{Kumara:2026uwi}, by solving the Einstein's equation in the presence of quantum corrected energy-momentum tensor, introduced by the $\kappa$-deformation parameter. The $\kappa$-deformed energy density as well the pressure components of this spherically symmetric energy-momentum tensor, is obtained by employing the $\kappa$-deformed Newton's potential in the standard Poisson's equation, in accordance with the standard conservation law. As a result, the explicit form of this $\kappa$-deformed Schwarzschild-AdS BH metric is given by 
\begin{equation}\label{a1}
 ds^2 = -f(r)dt^2 + f(r)^{-1}dr^2 +r^2d\Omega^2,
\end{equation}
where the spherically symmetric radial function $f(r)$ is
\begin{equation}\label{a2}
 f(r) = 1 - \frac{2M}{r} + \frac{a M^2}{2\pi r^3} - \frac{\Lambda r^2}{3},
\end{equation}
where $a$ is the $\kappa$-deformation parameter defined in the natural units \cite{Kumara:2026uwi}. The leading order correction introduces a $1/r^3$ term in the BH solution, in comparison with the $1/r^2$ and $1/r^4$ terms introduced by the leading corrections of Moyal space-time \cite{Nicolini:2005vd} and LQG \cite{Lewandowski:2022zce} frameworks. In all these cases, the NC and LQG corrections are seen to oppose the gravity, and consequently the higher order corrections of \cite{Nicolini:2005vd} removes the BH singularity through de-Sitter core. In this context, one would need to incorporate all the higher order $\kappa$-deformed corrections terms and employ non-perturbative approach to investigative whether the $\kappa$-deformation also leads to singularity resolution or not. 

The thermodynamic behaviour of BH can be studied in the extended phase space. For this, we first compute the mass of this BH, at the event horizon $r_+$, from the condition $f(r_+)=0$
\begin{equation}\label{a4}
 M = \frac{r_+}{2}\bigg( 1 + \frac{8\pi Pr_+^2}{3} \bigg) + \frac{a}{16\pi} \bigg( 1 + \frac{8\pi Pr_+^2}{3} \bigg)^2,
\end{equation}
where $P$ is the thermodynamic pressure. In general the temperature of a BH is defined in terms of the surface gravity as \(T=\frac{f'(r_+)}{4\pi}\). Substituting Eq.(\ref{a2}) in this definition, we get the explicit form of the Hawking temperature $T$ as
\begin{equation}\label{a6}
 T = \frac{1}{4\pi r_+} + 2Pr_+ - \frac{a}{16\pi^2r_+^2} \bigg( 1 + \frac{8\pi Pr_+^2}{3} \bigg)^2.
\end{equation}
Now the entropy of this BH can be defined in terms of the standard Bekenstein-Hawking area law as
\begin{equation}\label{a5}
 S = \frac{A}{4} = \pi r_+^2,
\end{equation}
However the entropy obtained from Hawking temperature (given in Eq.(\ref{a6})) through first law is inconsistent with the one defined in Eq.(\ref{a5}), via Bekenstein area law. Interestingly this inconsistency can be removed upon modifying the first law, where the modification has been shown to depend on the explicit mass
dependence of the energy-momentum tensor \cite{Ma:2014qma}. As a result the modified first law and associated Smarr relation for the $\kappa$-deformed Schwarzschild-AdS BH has been derived explicitly in \cite{Kumara:2026uwi}
\begin{equation}\label{a3}
  W\,dM = TdS + VdP + \Phi_a\,da,
\end{equation}
and
\begin{equation}\label{a10}
 WM = 2(TS - PV) + \Phi_a a,
\end{equation}
respectively. Here $W= 1-\frac{aM}{2\pi r_+^2}$ is the modification factor to the standard first law, derived by substituting Eq.(\ref{a2}) in the relation $W = 1 + \int_{r_+}^{\infty}\frac{\partial}{\partial r}\left(\frac{r}{2}\frac{\partial f}{\partial M}\right)dr$ (see \cite{Ma:2014qma} for complete derivation). As $W$ reduces to unity, one recovers the standard first law and Smarr relations of the commutative space-time. It is worth noting that $W dM$ present in Eq.(\ref{a3}) is not an exact differential and consequently affects in defining the differentials of the thermodynamic potentials.

Now we find that the temperature in Eq.(\ref{a6}) and entropy defined in Eq.(\ref{a5}) is consistent with the expression for the temperature from the modified first law
\begin{equation}\label{a7}
 T = W\bigg(\frac{\partial M}{\partial S}\bigg)_{P,a}.
\end{equation}
Similarly, the thermodynamic volume of this BH can be obtained from the modified first law as
\begin{equation}\label{a8}
 V = W\bigg(\frac{\partial M}{\partial P}\bigg)_{S,a} = \frac{4\pi r_+^3}{3}.
\end{equation}
The modified first law Eq.(\ref{a3}), has been constructed upon incorporating the $\kappa$-deformation parameter $a$ and its conjugate $\Phi_a$ as thermodynamic variables in the extended phase-space. Thus from Eq.(\ref{a3}), the conjugate potential is obtained as
\begin{equation}\label{a9}
 \Phi_a = W\bigg(\frac{\partial M}{\partial a}\bigg)_{S,P} = \frac{M^2}{4\pi r_+^2}.
\end{equation}
The equation of state for this $\kappa$-deformed Schwarzschild-AdS BH is given by
\begin{equation}\label{a11}
 T = \frac{1}{2\pi v} + Pv - \frac{a}{4\pi^2 v^2} \bigg( 1 + \frac{2\pi Pv^2}{3} \bigg)^2,
\end{equation}
where \(v=2r_+\) is the specific volume of BH \cite{Kubiznak:2012wp}. The critical behaviour of this BH is analyzed from the conditions $\big(\frac{\partial T}{\partial v}\big)_{a,P}=0,~\big(\frac{\partial^2 T}{\partial v^2}\big)_{a,P}=0$ and by solving these equations, we obtain the critical points and the associated critical ratio as
\begin{widetext}
\begin{equation}\label{a12}
\begin{split}
 v_c = \frac{2a(2^{1/3}+1)}{3\pi}, ~P_c = \frac{9\pi(2^{1/3}-1)}{8a^2}, ~T_c = \frac{3(2^{-1/3}+2^{1/3}-1)}{2a(1+2^{1/3})^2}.
\end{split}
\end{equation}
\end{widetext}
and
\begin{equation}\label{a13}
 \frac{P_c v_c}{T_c} = 0.370,
\end{equation}
respectively. This $\kappa$-induced critical ratio is very close to the Van der Waal's ratio \(0.375\) \cite{Kubiznak:2012wp} and is independent of the \(\kappa\)-deformation parameter \(a\). Similarly, the critical ratio of uncharged AdS BHs in Moyal space-time \cite{Wang:2024jlj} and LQG \cite{Wang:2024jtp} also do not depends on the minimal length scale, although the critical volume, temperature and pressure are deformed by QG corrections, showing the universality of critical ratio in the NC space-time and LQG frameworks. In this context, it is to be noted that the critical behaviour of the uncharged-AdS BHs in NC and LQG frameworks, are analogous to that of the Reissner-Nordstrom-AdS BH.

\section{Joule-Thomson expansion of $\kappa$-deformed BH}\label{sec3}

The Joule-Thomson expansion is an adiabatic isenthalpic throttling process in which a thermodynamic system undergoes a temperature change characterized by the Joule-Thomson coefficient $\mu$, whose sign determines whether the system undergoes cooling $(\mu>0)$ or heating $(\mu<0)$ during the expansion, such that the inversion curve separates these two regimes in the temperature-pressure plane. In this study, we investigate how the $\kappa$-deformation parameter induces the Joule-Thomson expansion behaviour in an uncharged Schwarzschild-AdS BH. We proceed this by calculating the Joule-Thomson coefficient associated with the $\kappa$-deformed Schwarzschild-AdS BH (discussed in Sec.(\ref{sec2})), and obtain the expression for the inversion temperature and pressure, at the inversion point. In order to begin this discussion, we first re-express the BH's temperature and pressure (of Sec.(\ref{sec2})) in terms of $M$, $r_+$ and $a$, respectively as 
\begin{equation}\label{d1}
\begin{split}
 P &= \frac{3 M}{4\pi r_+^3} - \frac{3}{8 \pi r_+^2} - \frac{3aM^2}{16\pi^2 r_+^5},\\
 T &= \frac{3 M}{2\pi r_+^2} - \frac{1}{2\pi r_+} - \frac{5aM^2}{8\pi^2r_+^4}.
\end{split}
\end{equation}
In general, the Joule-Thomson effect is an isenthalpic process, and since the BH mass $M$ is treated as the enthalpy, the Joule-Thomson expansion for AdS BHs corresponds to a constant mass process where $dM = 0$. The Joule-Thomson coefficient $\mu$ for the above discussed BH is defined as $\mu = \Big(\frac{\partial T}{\partial P}\Big)_{M,a}$. By using the expressions of Eq.(\ref{d1}) in this definition, we get the explicit form of Joule-Thomson coefficient for the $\kappa$-deformed Schwarzschild-AdS BH
\begin{equation}\label{d2}
 \mu  = \frac{8r_+\big( \pi r_+^3 - 6M \pi r_+^2  + 5 M^2 a \big)}{3\big( 4 \pi r_+^3 - 12 M \pi r_+^2  + 5 M^2 a\big)}.
\end{equation}

\begin{figure*}[!]
    \centering
    \begin{subfigure}{0.47\textwidth}
        \centering
        \includegraphics[width=\linewidth]{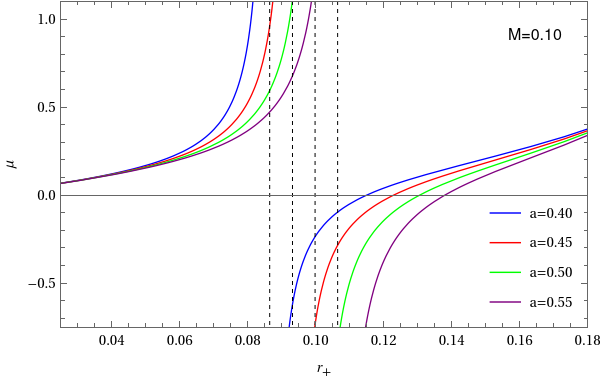}
        \caption{}
        \label{fig:plot1}
    \end{subfigure}
    \hfill
    \begin{subfigure}{0.47\textwidth}
        \centering
        \includegraphics[width=\linewidth]{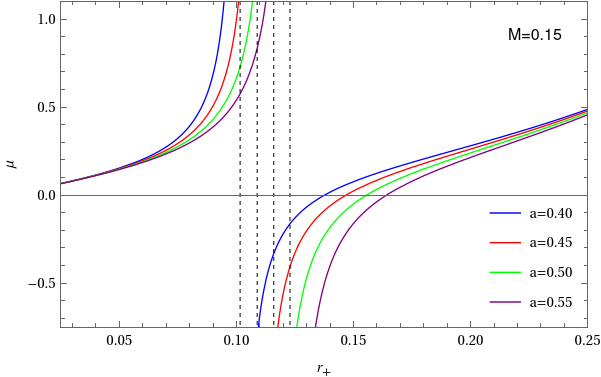}
        \caption{}
        \label{fig:plot2}
    \end{subfigure}
    
    \begin{subfigure}{0.47\textwidth}
        \centering
        \includegraphics[width=\linewidth]{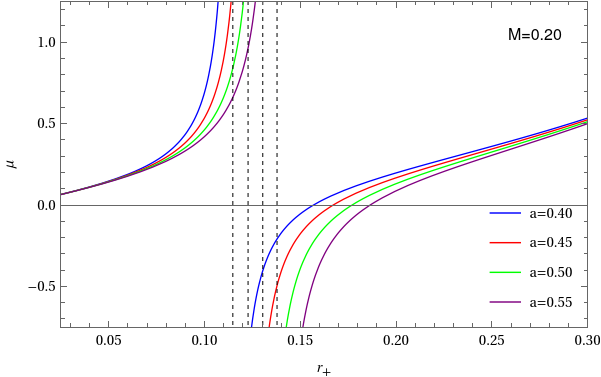}
        \caption{}
        \label{fig:plot3}
    \end{subfigure}
    \hfill
    \begin{subfigure}{0.47\textwidth}
        \centering
        \includegraphics[width=\linewidth]{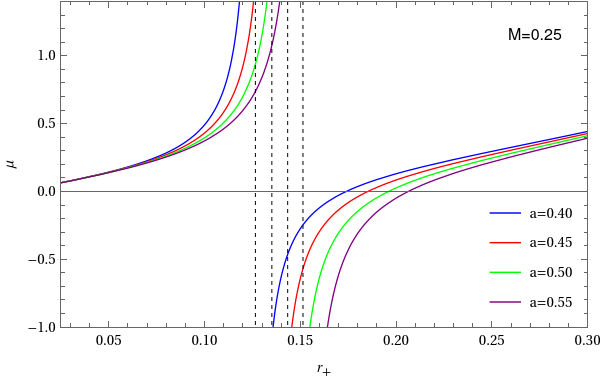}
        \caption{}
        \label{fig:plot4}
    \end{subfigure}
 
    \caption{Plot of $\mu$ against $r_+$ for $\kappa$-deformed Schwarzschild-AdS BH with (a) $M=0.10$, (b) $M=0.15$, (c) $M=0.20$ and (d) $M=0.25$ for different $a$ values}
\label{fig:jt}
\end{figure*}

Fig.(\ref{fig:jt}) shows the variation of $\mu$ along $r_+$ for fixed $M$ and $a$ values. As we move outward along increasing $r_+$, the $\mu$ also increases monotonically and diverges at certain $r_+$, as shown by the dashed vertical lines. This corresponds to the cooling process, which is described by the regions with $\mu>0$. We notice that for fixed $M$ values, the BHs with lesser $a$ values are seen to cool faster. Once the cooling process stops we observe that the sign of $\mu$ flips. When the $r_+$ increases further, $\mu<0$ and this corresponds to a heating process. Here the BHs with lesser $a$ values are seen to become hotter first. This heating process continues until it becomes $\mu=0$, which is known as the inversion point. 

At the inversion point $\mu=0$, we obtain the inversion temperature and pressure as
\begin{equation}\label{d3}
\begin{split}
 P_i &= \frac{9072 \pi ^2 r_+^2-6120 \pi  r_+ a-175 a^2}{14400 \pi ^2 r_+^3 a}\\
 T_i &= \frac{1296 \pi ^2 r_+^2-720 \pi  r_+ a-25 a^2}{1440 \pi ^2 r_+^2 a}
\end{split}
\end{equation}

\begin{figure}[!]
\includegraphics[width=0.5\textwidth]{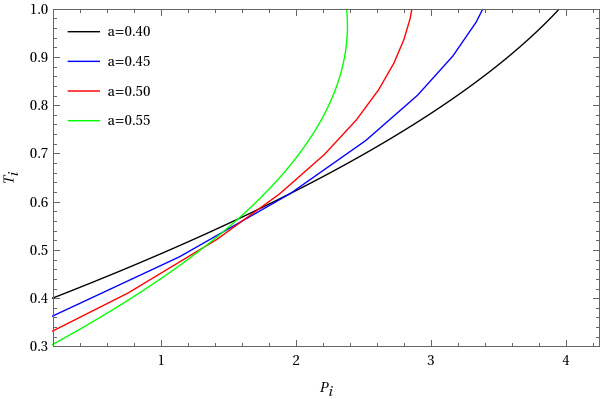}
\caption{Plot of $T_i$ against $P_i$ of the $\kappa$-deformed Schwarzschild-AdS for different $a$ values} \label{fig:tipi}
\end{figure} 

Fig.(\ref{fig:tipi}), show the inversion curves for different values of $a$. From this plot we notice that initially at lower pressure (larger $r_+$ values), the $1/a$ term of $T_i$ becomes dominant and in this regime the inversion temperature increases for lower $a$ values. But as the pressure increases further, the $1/r_+$ and $1/r_+^2$ terms of $T_i$ become significant, outweighing the $1/a$ suppression. As a result, $T_i$ increases with increasing $a$ values at the high pressure regime (lower $r_+$). Here the upper left region of the inversion curve corresponds to the cooling process and the lower right region corresponds to the heating process. In the commutative limit ($a\to 0$), both $T_i$ and $P_i$ diverges, and as a result the entire inversion curve recedes to infinity.  Thus we cannot observe the inversion curve anywhere in the finite physical $T_i-P_i$ region in the commutative limit. Therefore $\kappa$-deformation parameter is essential for introducing the inversion curve and Joule-Thomson expansion behaviour.

When $P_i=0$, we get the minimum inversion temperature from Eq.(\ref{d3}) as
\begin{equation}\label{d4}
 T_i^{(min)} = \frac{27}{5a(13 \sqrt{2} +17)}
\end{equation}
This minimum inversion temperature completely depends on the $\kappa$-deformation parameter just like the critical temperature, as discussed in above section. From Eq.(\ref{a13}) and Eq.(\ref{d4}), the ratio of minimum inversion temperature to critical temperature is estimated as
\begin{equation}\label{d5}
 \frac{T_i^{(min)}}{T_c} = 0.49316
\end{equation}

\begin{table*}[!]
\centering
\begin{tabular}{c @{\hspace{0.6cm}} c}
\hline\\
{BH solutions} & ${T_i^{(min)}}/{T_c}$ \\[0.2cm]
\hline\\
$\kappa$-Schwarzschild-AdS & $0.49316$ \\[0.2cm]
Reissner-Nordstom-AdS \cite{Kubiznak:2012wp} & $0.50000$\\[0.2cm]
Moyal-Schwarzschild-AdS \cite{Wang:2024jlj} & $0.58294$ \\[0.2cm]
LQG-Schwarzschild-AdS \cite{Wang:2024jtp} & $0.65205$ \\[0.2cm]
\hline
\end{tabular}
\caption{\small ${T_i^{(min)}}/{T_c}$ values for different BH solutions}
\label{tab:values}
\end{table*}

Eventhough $T_i^{(min)}$ and $T_c$, explicitly depends on the value of $a$, their ratio is independent of the $\kappa$-deformation parameter. Similarly, the ${T_i^{(min)}}/{T_c}$ ratio for uncharged AdS black holes within the frameworks of Moyal space-time \cite{Wang:2024jlj} and LQG \cite{Wang:2024jtp}, where the ratio ${T_i^{(min)}}/{T_c}$ also remains independent of the NC and LQG parameters (see Table.(\ref{tab:values})). We also find that our value of ${T_i^{(min)}}/{T_c}$ is in close agreement with value $0.5$ obtained for the standard Reissner-Nordstrom BH \cite{Kubiznak:2012wp}, since the $\kappa$-deformation parameter $a$ plays the role analogous to electric charge of RN-AdS BH, and the slight difference in their values arise from the difference in the powers of RN $(1/r^2)$ and $\kappa$-deformed correction $(1/r^3)$ terms. Interestingly, the thermodynamic behavior of the NC black hole is shown to be analogous to that of the Reissner-Nordstrom BH and the Moyal NC parameter has been shown to be related with the squared electric charge \cite{Kim:2008vi}.

\begin{figure*}[!]
    \centering
    \begin{subfigure}{0.47\textwidth}
        \centering
        \includegraphics[width=\linewidth]{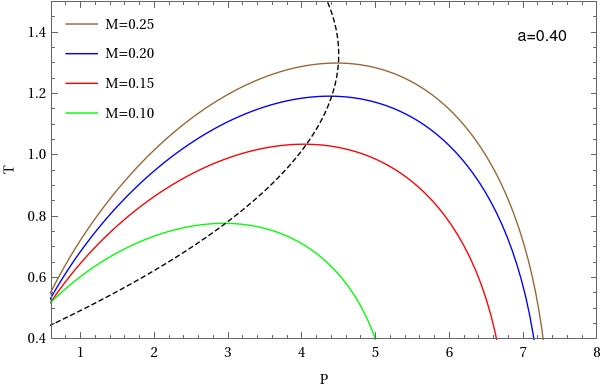}
        \caption{}
        \label{fig:plota}
    \end{subfigure}
    \hfill
    \begin{subfigure}{0.47\textwidth}
        \centering
        \includegraphics[width=\linewidth]{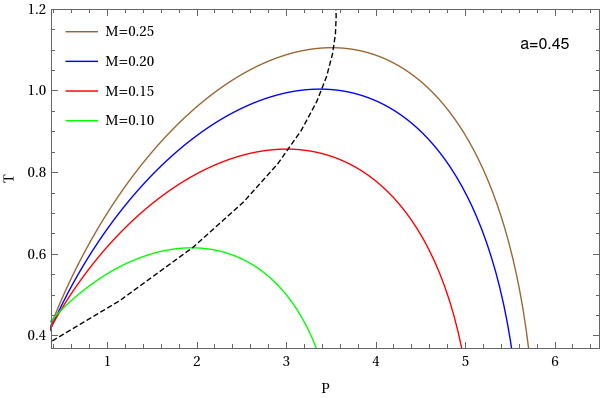}
        \caption{}
        \label{fig:plotb}
    \end{subfigure}
    
    \begin{subfigure}{0.47\textwidth}
        \centering
        \includegraphics[width=\linewidth]{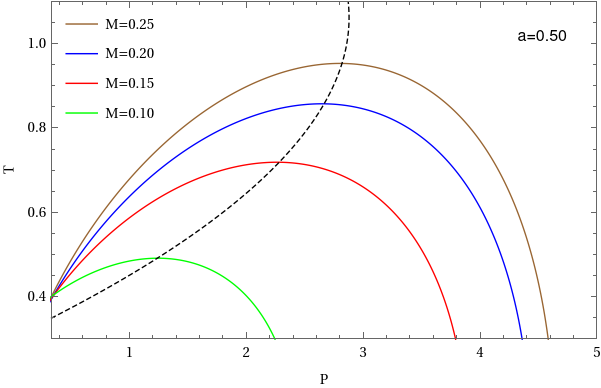}
        \caption{}
        \label{fig:plotc}
    \end{subfigure}
    \hfill
    \begin{subfigure}{0.47\textwidth}
        \centering
        \includegraphics[width=\linewidth]{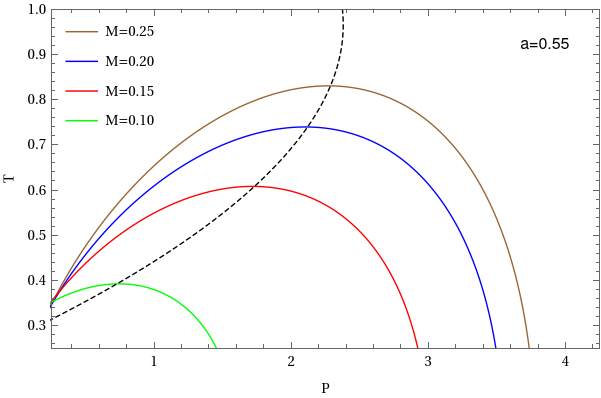}
        \caption{}
        \label{fig:plotd}
    \end{subfigure}
    
    \caption{Plot of $P$ against $T$ for $\kappa$-deformed Schwarzschild-AdS BH with (a) $a=0.40$, (b) $a=0.45$, (c) $a=0.50$ and (d) $a=0.55$ for the BH mass values $M=0.10,~0.15,~0.20,~0.25$. Here solid coloured lines represent isenthalpic curves and dotted line denotes the inversion curve.}
    \label{fig:pt}
\end{figure*}

Although the above defined $\kappa$-AdS BH exhibit Joule-Thomson effect, it is to be noted that the existence of the inversion point depends on the value of BH mass and hence one cannot observe this for an arbitrary value of $M$. Thus there exist a minimum mass value beyond which the inversion point do not exist. Incorporating the conditions $r_+>0,P>0,M>0$ at the inversion point where $\mu=0$, we obtain this minimum inversion mass $M_i^{(min)}$ as
\begin{equation}\label{m-min}
 M_i^{(min)}=\frac{5a}{32\pi}.
\end{equation}
Fig.(\ref{fig:pt}) shows the isenthalpic curves of the $\kappa$-deformed Schwarzschild-AdS BH, where the coloured lines represent the constant mass curves corresponding to the values $M=0.10,~0.15,~0.20,~0.25$ and dashed line represents the inversion curve. The regions left to the inversion curve corresponds to heating region where $\mu<0$ and region right to the inversion curve corresponds to cooling region where $\mu>0$. The point of intersection of inversion curve with the isenthalpic curve corresponds to the inversion point where $\mu=0$. However, the $\kappa$-deformed space-time non-commutativity introduces a minimum inversion mass, as shown in Eq.(\ref{m-min}). For \(M > M_i^{(\min)}\), the inversion condition \(\mu = 0\) yields a physical inversion point with \(P_i > 0\), and the inversion curve exists in the physical \(T\)–\(P\) plane of thermodynamic phase space, separating the cooling region (\(\mu > 0\)) and the heating region (\(\mu < 0\)), as shown in Fig.(\ref{fig:pt}). As \(M \to M_i^{(\min)}\), the inversion point moves toward \(P_i \to 0\), and the inversion temperature approaches \(T_i^{(\min)}\), where the cooling region shrinks and the Joule-Thomson expansion becomes vanishingly weak. At \(M = M_i^{(\min)}\), the inversion curve reaches \(P_i = 0\), and the cooling region disappears. For \(M < M_i^{(\min)}\), the pressure would become negative, and the inversion point lies outside the physical regime. Therefore no physical inversion curve exists in the \(T\)–\(P\) plane, and every isenthalpic curve lies entirely in the heating region (\(\mu < 0\)), where the BH always heats up during throttling. Consequently the Joule-Thomson expansion is absent in $M<M_i^{(min)}$ region. Thus, the boundary \(M = M_i^{(\min)}\) is the inversion point where \(P_i = 0\) and, for smaller masses (i.e., $M<M_i^{(min)}$), the $P_i$ is pushed to unphysical negative pressure, so that the Joule-Thomson expansion ceases to exist. Therefore it is a feature of the \(\kappa\)-deformed BH solution, where the Joule-Thomson effect exists only above a minimum mass set by the $\kappa$-deformation parameter, below which the inversion point is unphysical.

In \cite{Panja:2026hja}, the Joule-Thomson expansion exists even in the commutative limit, since the effect is fundamentally induced by the BH's electric charge, where the NC parameter enhances the Joule-Thomson coefficient and shift the inversion curves, while preserving the standard ratio ${T_i^{(min)}}/{T_c} = 0.5$. In contrast, our results demonstrate that the Joule-Thomson effect is entirely absent in the commutative limit and is induced solely by the NC geometry itself. This yields a different minimum inversion-to-critical temperature ratio as shown in Eq.(\ref{d5}), while also introducing a minimum inversion mass $M_i^{(min)}$ below which no inversion point exists. Thus, our work fundamentally establishes that NC geometry can act as the essential source of the Joule-Thomson effect, rather than merely modifying an existing charge-driven phenomenon.

\section{Conclusion}\label{sec4}

In this work, we have investigated the Joule-Thomson expansion behaviour in a \(\kappa\)-deformed Schwarzschild-AdS BH. Our study has revealed that the \(\kappa\)-deformation parameter induces the inversion temperature and leads to the Joule-Thomson expansion behavior in an uncharged Schwarzschild-AdS BH. Similarly the minimal length scale of Moyal NC space-time and LQG frameworks have also shown to introduce Joule-Thomson effect in uncharged Schwarzschild-AdS BH. These results demonstrate that the minimal length scale associated with NC parameters and LQG parameter is sufficient enough to generate the Joule Thomson effect in BHs, and hence these studies are crucial in understanding how the quantum gravitational corrections can manifest in BH thermodynamics.

Remarkably, we have shown that the ratio of minimum inversion temperature to the critical temperature, \(T_i^{(min)}/T_c = 0.49316\), is independent of the \(\kappa\)-deformation parameter, and is in close agreement with the standard value of $0.5$ obtained for the Reissner-Nordstrom BH. The $T_i^{(min)}/T_c$ ratio of quantum corrected-AdS BHs solutions, obtained in the Moyal space-time and LQG framework, have been also shown to be independent of the minimal length scales. Furthermore, we have shown the $\kappa$-deformation parameter introduces a minimum inversion mass \(M_i^{(min)}\), and as as result the Joule-Thomson expansion in $\kappa$-deformed BHs can only be observed when the mass of this BH exceeds this minimum value.

Our study suggest that the NC space-time geometries would a plausible framework for exploring quantum gravitational effects of BH thermodynamics. Further investigations upon incorporating higher-order corrections in the \(\kappa\)-deformation parameter, and extending this analysis to rotating \(\kappa\)-deformed BHs would deepen the understanding of the effects non-commutativity in BH thermodynamics.

\section*{Acknowledgment}

This work is supported by the National Natural Science Foundation of China (Grant No.~12275080), and the Innovative Research Group of Hunan Province (Grant No.~2024JJ1006).



\bibliography{reference}

\end{document}